\documentclass[aps,prl,twocolumn,reprint,floatfix,superscriptaddress,showpacs]{revtex4}
\usepackage{graphicx}
\usepackage[pdftex,colorlinks,citecolor=blue,bookmarks]{hyperref}
\usepackage{bm}
\usepackage{longtable}
\date{\today}

\begin{document}

\title{Shape and pairing fluctuations effects on neutrinoless double beta decay nuclear matrix elements}

\author{Nuria L\'opez Vaquero}
\affiliation{Departamento de F\'isica Te\'orica, Universidad
  Aut\'onoma de Madrid, E-28049 Madrid, Spain}
\author{Tom\'as R. Rodr\'iguez}
\affiliation{Institut f\"ur Kernphysik, Technische Universit\"at Darmstadt, Schlossgartenstr. 2, D-64289 Darmstadt, Germany}
\author{J. Luis Egido}
\affiliation{Departamento de F\'isica Te\'orica, Universidad
  Aut\'onoma de Madrid, E-28049 Madrid, Spain}

\begin{abstract}
Nuclear matrix elements (NME) for the most promising candidates to detect neutrinoless double beta decay have been computed with energy density functional methods including deformation and pairing fluctuations explicitly on the same footing. The method preserves particle number and angular momentum symmetries and can be applied to any decay without additional fine tunings. The finite range density dependent Gogny force is used in the calculations. 
An increase of 10\%-40\% in the NME with respect to the ones found without the inclusion of pairing fluctuations is obtained, reducing the predicted half-lives of these isotopes. 
\end{abstract}

\maketitle

The possible detection of lepton number violating processes such as neutrinoless double beta decay ($0\nu\beta\beta$) is one of the current main goals for particle and nuclear physics research. In this process, an atomic nucleus decays into its neighbor with two neutron less and two proton more emitting only two electrons. Fundamental questions about the nature of the neutrino such as its Dirac or Majorana character, its absolute mass scale as well as its mass hierarchy can be determined if this process is eventually measured~\cite{RMP_80_481_2008}. On the one hand, searching for $0\nu\beta\beta$ decays represents an extremely difficult experimental task because an ultra low background is required to distinguish the predicted scarce events from the noise. Recently, the controversial claim of detection in $^{76}$Ge by the Heidelberg-Moscow (HdM) collaboration~\cite{PLB_586_198_2004} has been overruled by the latest data released by EXO-200, KamLAND-Zen and GERDA collaborations~\cite{PRL_109_032505_2012,PRL_109_062502_2013,arXiv:1307.4720}. Nevertheless, these results are challenging the experiments that are already running or in an advanced stage of development to detect directly this process~\cite{EPJC_73_2330_2013,JPCS_381_012044_2012,JPCS_375_042018_2012,PPNP_64_267_2010, AIPCP_942_101_2007,PRC_80_032501_2009,JPCS_365_042023_2012,PRC_78_035502_2008,JINST_8_P04002_2013,PRL_109_032505_2012}. 
On the other hand, in the most probable electroweak mechanism to produce $0\nu\beta\beta$, namely, the exchange of light Majorana neutrinos~\cite{PPNP_12_409_1984,RMP_80_481_2008}, the half-life of this process is inversely proportional to the effective Majorana neutrino mass $\langle m_{\nu}\rangle$, a kinematic phase space factor $G_{01}$ and the nuclear matrix elements $M^{0\nu}$ (NME):
\begin{equation}
\left[T^{0\nu}_{1/2}(0^{+}\rightarrow 0^{+})\right]^{-1}=G_{01}\left|M^{0\nu}\right|^{2}\left(\frac{\langle m_{\nu}\rangle}{m_{e}}\right)^{2}
\label{M_op}
\end{equation} 
where $m_{e}$ is the electron mass and $\langle m_{\nu}\rangle=|\sum_{k}U^{2}_{ek}m_{k}|$ is the combination of the neutrino masses $m_{k}$ provided by the neutrino mixing matrix $U$. The kinematic phase space factor can be determined precisely from the charge, mass and the energy available in the decay~\cite{PRC_85_034316_2012} while the nuclear matrix elements must be calculated using nuclear structure methods. The most commonly used ones are the quasiparticle random phase approximation~\cite{PRC_60_055502_1999,PRC_77_045503_2008,PRC_83_034320_2011,PRC_75_051303_2007,PRC_87_064302_2013} (QRPA), large scale shell model~\cite{PRL_100_052503_2008,NPA_818_139_2009,PRC_86_067304_2012} (LSSM), interacting boson model~\cite{PRC_79_044301_2009,PRC_87_014315_2013} (IBM), projected Hartree-Fock-Bogoliubov~\cite{PRC_82_064310_2010} (PHFB) and energy density functional~\cite{PRL_105_252503_2010,PPNP_66_436_2011,PLB_719_174_2013} (EDF). In recent years, most of the basic nuclear structure aspects of the NMEs have been understood within these different frameworks. In particular, the decay is favored when the initial and final nuclear states have similar intrinsic deformation~\cite{JM,PRL_105_252503_2010,PLB_719_174_2013}.  Indications~\cite{PRC_77_045503_2008,PRC_87_064302_2013,NPA_818_139_2009,PRL_105_252503_2010,PLB_719_174_2013} about the strong sensitivity of the transition operator to pairing correlations
suggest that fluctuations in this degree of freedom will play a relevant role in the description of this process. The
purpose of this  Letter is to report the first calculations of $0\nu\beta\beta$ NMEs including self-consistently shape and pairing fluctuations on the same footing within the EDF method. The finite range of the interaction used in the calculations (Gogny~\cite{NPA_428_23_1984}), with a common source for the long and short range parts of the force, guarantees a self-consistent interplay of  the shape and pairing fluctuations.
In this framework, following the generator coordinate method (GCM)~\cite{RingSchuck,RMP_75_121_2003}, the many body nuclear states are described as a linear combination (mixing) of particle number and angular momentum projected Hartree-Fock-Bogoliubov (HFB) wave functions with different shapes and pairing content~\cite{PLB_704_520_2011}:
\begin{figure}[t]
\begin{center}
\includegraphics [width=1\columnwidth]{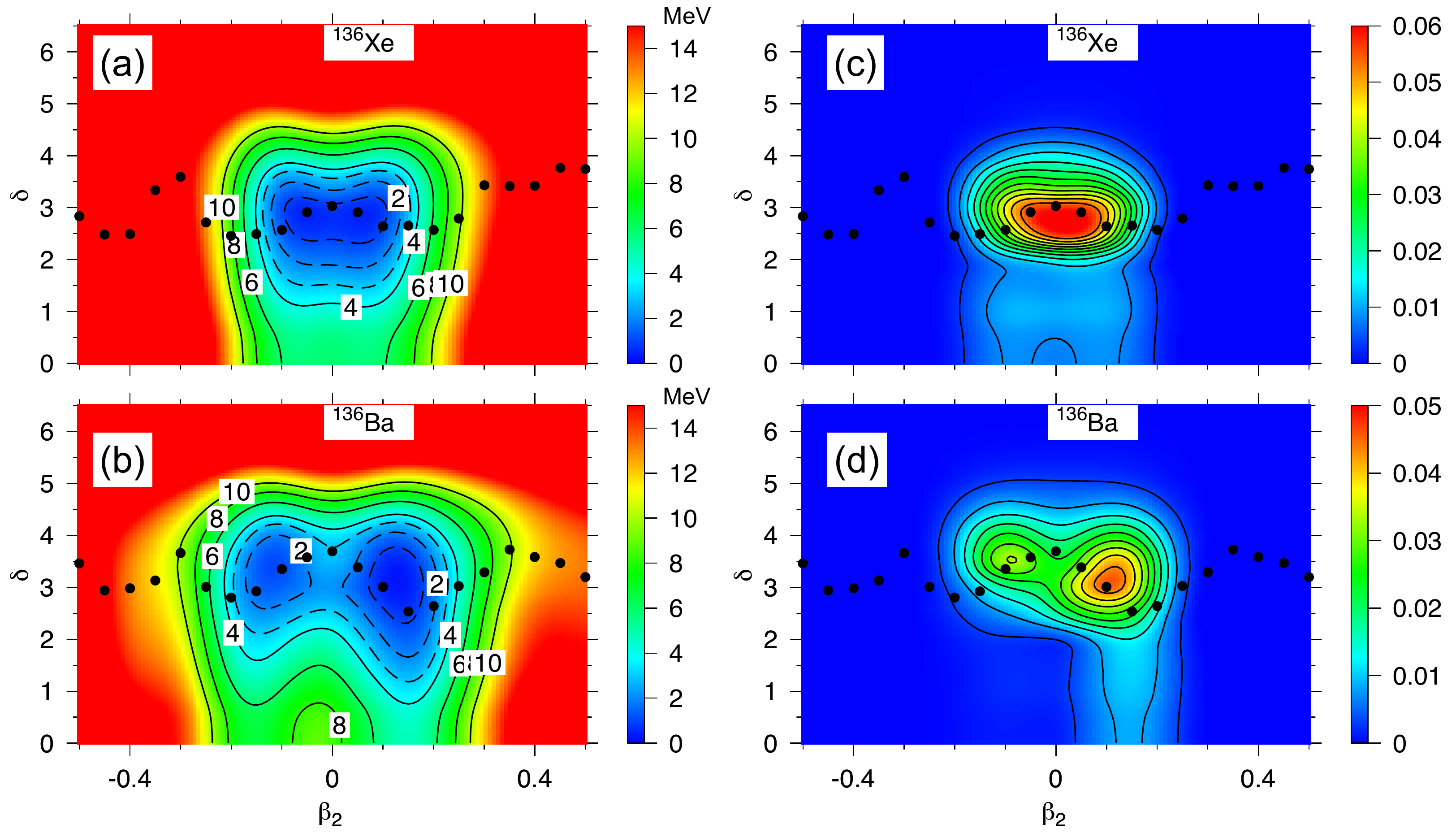}
\end{center}
\caption{(color online) Left: Particle number and angular momentum $I=0$ projected potential energy surfaces: $E^{I=0}(\beta_{2},\delta)=\langle\Psi^{I=0}(\beta_{2},\delta)|\hat{H}|\Psi^{I=0}(\beta_{2},\delta)\rangle/\langle\Psi^{I=0}(\beta_{2},\delta)|\Psi^{I=0}(\beta_{2},\delta)\rangle$ for (a) $^{136}$Xe and (b) $^{136}$Ba. Dashed and continuous lines are separated 1 MeV and 2 MeV respectively. The curves are normalized to their corresponding absolute minima. Right: Collective wave functions squared -$|G^{I=0;\sigma=1}(\beta_{2},\delta)|^{2}=|\sum_{\beta'_{2},\delta'}\langle\Psi^{I=0}(\beta_{2},\delta)|\Psi^{I=0}(\beta'_{2},\delta')\rangle^{1/2}$ $g^{I=0;\sigma=1}(\beta'_{2},\delta')|^{2}$- for (c) $^{136}$Xe and (d) $^{136}$Ba. The dots indicate the values of $\delta$ obtained in a self-consistent one dimensional calculation along $\beta_{2}$.}
\label{Fig1}
\end{figure}
\begin{equation}
|I^{+\sigma}_{i/f}\rangle=\sum_{\beta_{2},\delta}g^{I\sigma}_{i/f}(\beta_{2},\delta)|\Psi^{I}_{i/f}(\beta_{2},\delta)\rangle
\label{GCM_WF}
\end{equation}
where $I$ is the angular momentum, $\sigma$ labels the different states for a given angular momentum, $\beta_{2}$ and $\delta$ are the intrinsic axial quadrupole and pairing degrees of freedom respectively, $g^{I\sigma}_{i/f}(\beta_{2},\delta)$ are the coefficients found by solving the Hill-Wheeler-Griffin (HWG) equations~\cite{RingSchuck,PLB_704_520_2011} and the projected wave functions are defined as:
\begin{equation}
|\Psi^{I}_{i/f}(\beta_{2},\delta)\rangle=P^{N_{i/f}}P^{Z_{i/f}}P^{I}|\phi(\beta_{2},\delta)\rangle
\label{PROJ_WF}
\end{equation}
with $P^{N(Z)}$ and $P^{I}$ being the neutron (proton) number and angular momentum projection operators respectively. Shape and pairing degrees of freedom are included on the same footing through the different HFB-type states $|\phi(\beta_{2},\delta)\rangle\equiv|\phi\rangle$. These wave functions are found by minimizing the particle number projected energy -variation after projection (PN-VAP) method~\cite{NPA_696_467_2001}- with constraints both in the mean value of the axial quadrupole moment operator $\langle\phi|\hat{Q}_{20}|\phi\rangle=\frac{\beta_{2}3r_{0}^{2}A^{5/3}}{\sqrt{20\pi}}$ and in the particle number fluctuations $\langle \phi|(\Delta\hat{A})^2|\phi \rangle^{1/2} = \delta$~\cite{footnote}, being $r_{0}=1.2$ fm and $A$ the mass number.
One of the benefits of the PN-VAP method with a constraint in $\delta$ is the proper treatment of pairing correlations and the absence of a pairing gap collapse found in the BCS or plain HFB methods in the weak pairing regime. Both the calculation of the intrinsic states and the HWG diagonalization are performed with the same underlying interaction, Gogny D1S~\cite{NPA_428_23_1984}. Once the HWG equations are solved, any observable such as energy spectra, radii, electromagnetic transitions, fission barriers, etc.~\cite{RMP_75_121_2003} and, more interestingly, $0\nu\beta\beta$
NMEs can be found within the same formalism (see Refs.~\cite{PRL_105_252503_2010,PPNP_66_436_2011,PLB_719_174_2013} and references therein for more details). To expand the HFB-like wave functions a large configuration space including eleven major harmonic oscillator shells is used and the number of such intrinsic states is up to 440 for each nucleus with $\beta_{2}\in[-0.85,0.95]$ and $\delta\in[0.5,6.5]$.

Particle number and rotational symmetry restorations are included within this framework as well as pairing, quadrupole deformation and quantum fluctuations of both collective degrees of freedom. However, triaxiality, octupolarity, isospin restoration or explicit quasiparticle excitations are missing in this approach and their influence on the NMEs (or any other observable) is beyond the scope of this work. 
Concerning the specific details about the NMEs, these quantities are computed as the sum of Fermi (F) and Gamow-Teller (GT)
terms~\cite{RMP_80_481_2008}(tensor contribution is neglected in this
work~\cite{NPA_818_139_2009,PRC_75_051303_2007}):
\begin{equation}
M^{0\nu}=-\left(\frac{g_{V}}{g_{A}}\right)^{2}M^{0\nu}_{F} +M^{0\nu}_{GT}
\label{MMM}
\end{equation}
with $g_{V}=1$ and $g_{A}=1.25$ being the vector and axial coupling
constants. In addition, the closure approximation~\cite{RMP_80_481_2008,PRC_83_015502_2011} is used due to the impossibility of calculating at the same level of accuracy the odd-odd intermediate nucleus. The neutrino potentials include finite size, higher order currents and short range correlations corrections and their parameters are the same as in Refs.~\cite{NPA_818_139_2009,PRL_105_252503_2010}. 
\begin{figure}[t]
\begin{center}
\includegraphics [width=1.0\columnwidth]{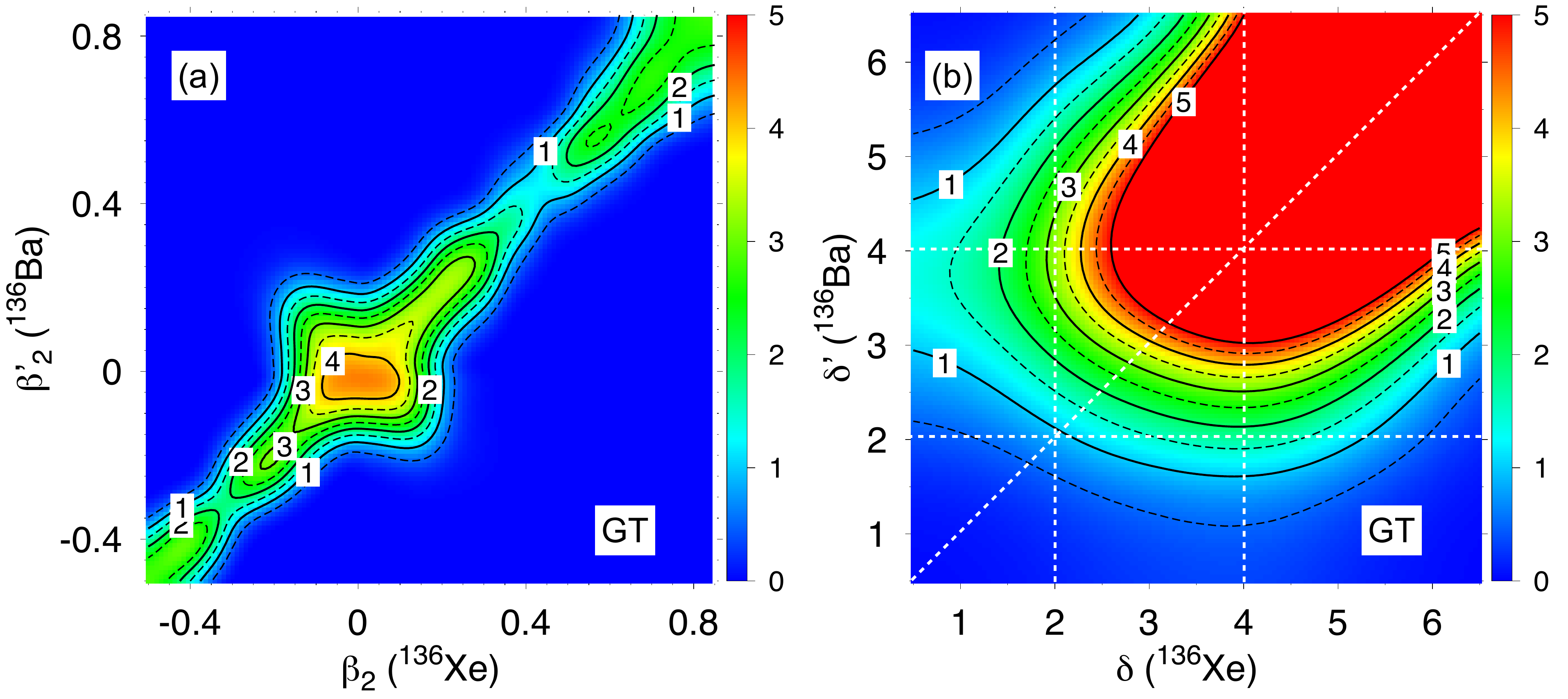}
\end{center}
\caption{(color online) Particle number and angular momentum $I=0$ projected GT-NMEs as function of (a) the quadrupole deformation and (b) pairing of the initial $^{136}$Xe and final $^{136}$Ba states. The horizontal and vertical dotted lines delimit the region where the wave functions of both nuclei  take the largest values. Contour lines are separated 0.5 units.}
\label{Fig3}
\end{figure}

We now discuss in detail the decay of the $^{136}$Xe $\rightarrow$ $^{136}$Ba to illustrate the method. The starting point is the determination of the mixing weights of the initial and final states (Eq.~\ref{GCM_WF}). To shed light on the physical insight of these states we analyze first the potential energy surfaces (PES) computed with the wave functions given in Eq.~\ref{PROJ_WF} with $I=0$ (see Fig.~\ref{Fig1}(a)-(b)). In $^{136}$Xe we obtain a rather symmetric PES around $\beta_{2}=0$, with two degenerated minima at $(\beta_{2}=\pm0.05,\delta=3)$. The energy increases significantly by increasing the deformation from $\beta_{2}=\pm0.15$ and also by enlarging the pairing content from $\delta\approx4$. On the other hand, a wider PES (both in $\beta_{2}$ and $\delta$) with two minima at ($\beta_{2}=0.15,\delta=3$) -the absolute one- and ($\beta_{2}=-0.10,\delta=3.5$) are obtained for $^{136}$Ba. The absolute minimum in this case is softer in the $\delta$ direction than the second one and the energy also rises considerably for $\beta_{2}>\pm0.2$ and $\delta>5$. More interestingly, the softness of the PESs at $\delta\in[1,4]$ in the interval of shapes ranging from $\beta_{2}\in[-0.2,0.2]$ is ignored in one dimensional calculations in the $\beta_{2}$ direction represented by the dots although this effect can play a role in the final structure of the states. 

This is confirmed by the ground state collective wave functions evaluated from the weights $g^{I\sigma}(\beta_2,\delta)$ and shown in Fig.~\ref{Fig1}(c)-(d). For $^{136}$Xe -Fig.~\ref{Fig1}(c)- a practically spherical distribution is found at the position of the potential wells represented in Fig.~\ref{Fig1}(a), as it should correspond to a semi magic nucleus $(N=82)$. For the $^{136}$Ba ground state -Fig.~\ref{Fig1}(d)- two maxima are obtained around the corresponding potential wells of Fig.~\ref{Fig1}(a), although the distribution is more concentrated in the prolate one. Nevertheless, we obtain large weights in the collective wave functions within an interval of $\delta\in[2,4]$ and this mixing is not taken into account in a 1D calculation.

Ground state observables can be directly computed within the present EDF method and compare with the experimental data (see Table~\ref{Table1}). In the nuclei discussed above, we obtain a very good agreement for the radii and a slight overestimation of the binding energies. The latter is a rather general result because the interaction was globally fitted at the mean field level and beyond mean field correlations will produce extra binding energy. Nevertheless, the differences from data is not larger than 5 MeV, which is within the precision of the Gogny D1S interaction for masses~\cite{EPJA_33_237_2007}. Finally, total Gamow-Teller strengths for initial $(S_{-})$ and final $(S_{+})$ states are also rather well reproduced assuming a quenching factor of $(0.74)^{2}$~\cite{RMP_77_427_2005,PPNP_66_436_2011}.

The dependence of the nuclear matrix elements on the collective variables $(\beta_{2},\delta)$ can be studied straightforwardly within the EDF method by computing the transition matrix elements between the projected states (Eq.~\ref{PROJ_WF}):
\begin{eqnarray}
M^{0\nu}_{F/GT}(\beta_{2},\delta;\beta'_{2},\delta')=\nonumber\\
\frac{\langle\Psi^{I=0}_{i}(\beta_{2},\delta)|\hat{M}^{0\nu}_{F/GT}|\Psi^{I=0}_{f}(\beta'_{2},\delta')\rangle}{\langle\Psi^{I=0}_{i}
(\beta_{2},\delta)|\Psi^{I=0}_{i}\beta_{2},\delta)\rangle^{1/2} \langle\Psi^{I=0}_{f}(\beta'_{2},\delta')|\Psi^{I=0}_{f}(\beta'_{2},\delta')\rangle^{1/2}}\nonumber\\
\label{NME_bd}
\end{eqnarray}

where $\hat{M}^{0\nu}_{F/GT}$ are two body operators including Fermi and Gamow-Teller neutrino potentials and spin and isospin dependences~\cite{PLB_719_174_2013}.
We now analyze separately the influence of the two degrees of freedom considered here on the NMEs and in the GT part (the Fermi part presents a similar behavior and it is not shown here). To do so, we fix first in Eq.~\ref{NME_bd} the values of $(\delta=\delta'=3)$ -chosen to be inside the relevant part in Fig.~\ref{Fig1}(c)-(d)- and represent the NME as a function of the quadrupole deformation of the initial and final states in Fig.~\ref{Fig3}(a). We obtain that the strength of the transition is larger when the decay is between similar deformations for the initial and final states -diagonal part of the Fig.~\ref{Fig3}(a). In addition, spherical shapes are also preferred and non-diagonal matrix elements have a significant value around this configuration $(\beta_{2}=-\beta'_{2})$. This behavior has been already reported in previous works within the EDF and LSSM frameworks~\cite{PRL_105_252503_2010,PPNP_66_436_2011,PLB_719_174_2013,JM}. On the other hand, we study the dependence of the NME on the pairing degree of freedom fixing the deformations of the initial and final states at the values where the maximum of the $^{136}$Ba collective wave function is found $(\beta_{2}=0.1)$ and leaving free the values for $(\delta,\delta')$ -see Fig.~\ref{Fig3}(b). 

Vanishing matrix elements are obtained for $\delta < 2$ and $\delta^{\prime} < 2$. However, for $\delta(\delta^{\prime})$ values  larger than $2$ the matrix element grows rapidly with increasing $\delta(\delta^{\prime})$ in the band region 
$\delta^{\prime} \approx \delta-3$ and $\delta^{\prime} \approx \delta+3$.  A correlation between pairing and NME has been also previously reported indirectly~\cite{PRL_105_252503_2010,PLB_719_174_2013,PRC_87_064302_2013} but it is explicitly shown for the first time in this work. Furthermore, the distribution is quite wide meaning that pairing mixing plays an important role.

\begin{table*}[t]
\caption{Columns (2-7): theoretical and experimental binding energies~\cite{CPC_36_1603_2012} (in MeV), radii~\cite{ADNDT_87_185_2004} (in fm) and total Gamow-Teller strength~\cite{PRL_103_012503_2009,PRC_40_540_1989,PRC_55_2802_1997,PRC_78_041602_2008,PRC_71_054313_2005} -$S_{-(+)}$ for the initial (final) state- for the $0\nu\beta\beta$ candidates. Theoretical values for $S_{-(+)}$ are quenched by a factor $(0.74)^{2}$ .Columns (8-9): nuclear matrix elements for the most probable $0\nu\beta\beta$ emitters considering shape fluctuations $(\beta_{2})$ and both shape and pairing fluctuations $(\beta_{2},\delta)$ explicitly. Superscript and subscript values correspond to the Gamow-Teller -$M^{0\nu}_{GT}$- and Fermi -$\left(\frac{-g_{V}}{g_{A}}\right)^{2}M^{0\nu}_{F}$- components respectively. The last two columns are the variation of the NME and half-lives when the additional pairing degree of freedom is included.}
\begin{center}
\begin{tabular}{c|c|c|c|c|c|c||c|c|c|c}\hline \hline
Isotope  & $(BE)^{th}$ & $(BE)^{exp}$ & $R^{th}$ & $R^{exp}$ & $S^{th}_{+/-}$ & $S^{exp}_{+/-}$ & $M^{0\nu}$($\beta_{2}$)  & $M^{0\nu}$($\beta_{2},\delta$) & Var (\%) & $\frac{T_{1/2}(\beta_{2},\delta)}{T_{1/2}(\beta_{2})}$\\ \hline \hline 
$^{48}$Ca & 420.919 & 415.991 & 3.467 & 3.473 & 13.48 & $14.4\pm 2.2 $ & $2.370^{1.914}_{0.456}$ & $2.229^{1.797}_{0.431}$ & -6 & 1.13\\
$^{48}$Ti & 423.753 & 418.699 & 3.560 & 3.591 & 1.94 & $ 1.9\pm0.5 $ & & & & \\\hline
$^{76}$Ge & 664.604 & 661.598 & 4.025 & 4.081 & 20.96 & 19.89 & $4.601^{3.715}_{0.886}$ & $5.551^{4.470}_{1.082}$ & 21& 0.69 \\
$^{76}$Se & 665.268 & 662.072 & 4.075 & 4.139 & 1.26 & $1.45\pm0.07$& & & & \\\hline
$^{82}$Se & 717.034 & 712.842 & 4.122 & 4.139 & 23.57 & $21.91$ & $4.218^{3.381}_{0.837}$ & $4.674^{3.743}_{0.931}$ & 11 & 0.81 \\
$^{82}$Kr & 718.220 & 714.273 & 4.131 & 4.192 & 1.26 & & & & & \\\hline
$^{96}$Zr & 829.801 & 828.995 & 4.298 & 4.349 & 27.73 & & $5.650^{4.618}_{1.032}$ & $6.498^{5.296}_{1.202}$ & 15 & 0.76 \\
$^{96}$Mo & 834.212 & 830.778 & 4.320 & 4.384 & 2.64 & $ 0.29\pm0.08 $ & & & & \\\hline
$^{100}$Mo & 862.003 & 860.457 & 4.373 & 4.445 & 28.04 & 26.69 & $5.084^{4.149}_{0.935}$ & $6.588^{5.361}_{1.227}$ & 30 & 0.60 \\
$^{100}$Ru & 865.230 & 861.927 & 4.388 & 4.453 & 2.63 & & & & & \\\hline
$^{116}$Cd & 988.809 & 987.440 & 4.567 & 4.628 & 34.40 & 32.70 & $4.795^{3.931}_{0.864}$ & $5.348^{4.372}_{0.976}$ & 12 & 0.80 \\
$^{116}$Sn & 991.390 & 988.684 & 4.569 & 4.626 & 2.61 & $1.09\pm0.13$ & & & & \\\hline
$^{124}$Sn & 1051.981 & 1049.96 & 4.622 & 4.675 & 40.71 & & $4.808^{3.893}_{0.916}$ & $5.787^{4.680}_{1.107}$ & 20 & 0.69 \\
$^{124}$Te & 1052.019 & 1050.69 & 4.664 & 4.717 & 1.63 & & & & & \\\hline
$^{128}$Te  & 1082.541 & 1081.44 & 4.685 & 4.735 & 40.48 & 40.08 & $4.107^{3.079}_{1.027}$ & $5.687^{4.255}_{1.432}$ & 38 & 0.52 \\
$^{128}$Xe  & 1081.249 & 1080.74 & 4.724 & 4.775 & 1.45 & & & & & \\\hline
$^{130}$Te & 1097.320 & 1095.94 & 4.695 & 4.742 & 43.69 & 45.90 & $5.130^{4.141}_{0.989}$ & $6.405^{5.161}_{1.244}$ & 25 & 0.64 \\
$^{130}$Xe & 1097.655 & 1096.91 & 4.733 & 4.783 & 1.33 & & & & & \\\hline
$^{136}$Xe & 1143.500 & 1141.88 & 4.757 & 4.799 & 46.77 & & $4.199^{3.673}_{0.526}$ & $4.773^{4.170}_{0.604}$ & 14 & 0.77 \\
$^{136}$Ba & 1143.606 & 1142.77 & 4.789 & 4.832 & 1.06 & & & & & \\\hline
$^{150}$Nd & 1234.729 & 1237.45 & 5.033 & 5.041 & 50.35 & & $1.707^{1.278}_{0.429}$ & $2.190^{1.639}_{0.551}$ & 29 & 0.61 \\
$^{150}$Sm & 1236.249 & 1239.25 & 4.987 & 5.040 & 1.54 & & & & & \\
\hline \hline
\end{tabular}
\end{center}
\label{Table1}
\end{table*}%

The final step in the calculation of the NME is to consider the shape and pairing fluctuations present in the initial and final wave functions (Fig.~\ref{Fig1}(c)-(d)). Taking into account the wave function shapes and looking at Fig.~\ref{Fig3}(b) we find that the relevant part is the square defined by the intersection of the horizontal and vertical lines. Here we see that the pairing fluctuations allow a large richness of values of the nuclear matrix element (from zero up to approximately 5) which definitively contribute to the final value.

The results for the most probable candidates to detect $0\nu\beta\beta$ decays are summarized in Table~\ref{Table1}. We find in the $^{136}$Xe decay discussed above a 14\% larger NME when the pairing degree of freedom is explicitly included which leads to a reduction of the half-life in a factor 0.77. This result is consistent with exploring regions with larger values of the NME in the pairing degree of freedom thanks to the fluctuations in $\delta$ included in the collective wave functions. The same effect happens for the rest of candidates where the NME obtained including both deformation and pairing fluctuations are increased from 10\% to 40\% with respect to the values found by considering only shape mixings. The $^{48}$Ca is the only particular case where, due to its double magic character, the initial wave function is significantly moved towards less pairing correlations, thus giving a slightly smaller NME. 
Except for this decay, the updated NMEs lead to a reduction of the predicted half-lives up to factors from 0.81 ($^{82}$Se) to 0.52 ($^{128}$Te). Furthermore, a shorter $^{76}$Ge half-life as a function of the $^{136}$Xe one is predicted in the region allowed by HdM, IGEX~\cite{PRD_65_092007_2002}, GERDA, EXO-200 and KamLAND-Zen experiments. However, the HdM claim is incompatible both with the previous and these new values of the NMEs. 

Compared to other methods the new NMEs are getting closer to QRPA/IBM results for $^{48}$Ca, $^{76}$Ge, $^{128}$Te and $^{150}$Nd decays while they are the largest ones for the other candidates -see Fig. 7 of Ref.~\cite{PRC_87_014315_2013} for updated values. However, neither QRPA nor IBM calculations have explored explicitly this degree of freedom so far. On the other hand, these values move away from the LSSM ones and some work is in progress to study the NMEs along isotopic chains to disentangle the similarities/differences between both methods~\cite{inprep,PLB_719_174_2013}.

Part of this disagreement could be produced by the large values of the Fermi part obtained within QRPA, IBM and EDF methods compared to the LSSM ones that has been recently discussed in terms of isospin symmetry violation. Hence, spurious contributions to Fermi -and possibly GT- matrix elements exist in those cases where the initial and final states are not isospin eigenstates. In Ref.~\cite{PRC_87_045501_2013} is shown in the QRPA framework that correcting the parameters to have the Fermi part of the $2\nu\beta\beta$ decay equal to zero, the $M^{0\nu}_{F}$ is reduced but $M^{0\nu}_{GT}$ is barely affected. In Table~\ref{Table1} we show separately the GT and F components of the NME and we see that the gain including pairing fluctuations is similar in both channels. This fact could indicate that the observed increase is not produced by a stronger isospin symmetry violation. 

In summary, we have presented calculations for 0$\nu\beta\beta$ matrix elements within the EDF framework, including for the first time pairing and quadrupole axial deformation fluctuations together. We have confirmed  that  NMEs between states with similar quadrupole deformation are largest. Concerning the pairing degree of freedom  we found the following characteristics of the the NMEs: 1.- They are zero for weakly correlated states,  $\delta$ and $\delta^\prime<2$, 2.- They grow considerably for increasing pairing correlations and 3.- There exists a set of states belonging to a band along the main diagonal, defined by $\delta{^\prime} =\pm 3$, with large NMEs. 
This effect and the allowance of having pairing fluctuations in the initial and final wave functions produce a rise in the NMEs from 10\% to 40\% with respect to the values obtained without including them. The updated values reduces correspondingly the expected half-lives for the most probable candidates. 

T.R.R. thanks G. Mart\'inez-Pinedo for fruitful discussions. This work was partly supported from the Spanish Ministerio de  Ciencia e Innovaci\'on under contract  FPA2011-29854-C04-04 and by the Spanish Consolider-Ingenio 2010 Programme CPAN (CSD2007-00042). N.L.V acknowledges a scholarship of the Programa de Formaci\'on de Personal  Investigador (Ref. BES-2010-033107). T.R.R. acknowledges support from BMBF-Verbundforschungsprojekt number 06DA7047I and Helmholtz International Center for FAIR program.


\begin{thebibliography}{9}   
\bibitem{RMP_80_481_2008}
F.~T. Avignone, S.~R. Elliot, J. Engel, Rev. Mod. Phys. 80, 481 (2008).
\bibitem{PLB_586_198_2004} 
H. V. Klapdor-Kleingrothaus \textit{et al.}, Phys. Lett. B 586, 198 (2004).
\bibitem{PRL_109_032505_2012}
M. Auger et al., Phys. Rev. Lett. 109, 032505 (2012).
\bibitem{PRL_109_062502_2013}
A. Gando \textit{et al.}, Phys. Rev. Lett. 110, 062502 (2013).
\bibitem{arXiv:1307.4720}
M. Agostini \textit{et al.},  arXiv:1307.4720 (2013).
\bibitem{EPJC_73_2330_2013}
K.-H. Ackermann \textit{et al.},  Eur. Phys. J. C 73, 2330 (2013).
\bibitem{JPCS_381_012044_2012}
D. G. Phillips II \textit{et al.}, J. Phys.: Conf. Ser. 381 012044 (2012).
\bibitem{JPCS_375_042018_2012}
I. Ogawa \textit{et al.}, J. Phys.: Conf. Ser. 375 042018 (2012).
\bibitem{PPNP_64_267_2010}
K. Zuber, Prog. Part .Nucl. Phys. 64, 267 (2010).
\bibitem{AIPCP_942_101_2007}
K. Zuber \textit{et al.}, AIP Conf. Proc. 942, 101 (2007).
\bibitem{PRC_80_032501_2009}
J. Argyriades \textit{et al.}, Phys. Rev. C 80, 032501(R) (2009).
\bibitem{JPCS_365_042023_2012}
H Bhang \textit{et al.}, J. Phys.: Conf. Ser. 375, 042023 (2012).
\bibitem{PRC_78_035502_2008}
C. Arnaboldi \textit{et al.}, Phys. Rev. C 78, 035502 (2008).
\bibitem{JINST_8_P04002_2013}
V. \'Alvarez \textit{et al.}, JINST 8, P04002 (2013).
\bibitem{PPNP_12_409_1984}
W.C. Haxton and G.S. Stephenson, Prog. Part. Nucl. Phys. 12, 409 (1984).
\bibitem{PRC_85_034316_2012}
J. Kotila and F. Iachello, Phys. Rev. C 85, 034316 (2012).
\bibitem{PRC_60_055502_1999} 
F. \u{S}imkovic \textit{et al.}, Phys. Rev. C 60, 055502 (1999). 
\bibitem{PRC_77_045503_2008} 
F. \u{S}imkovic \textit{et al.}, Phys. Rev. C 77, 045503 (2008). 
\bibitem{PRC_83_034320_2011}
D.-L. Fang, A. Faessler, V. Rodin, and F. \u{S}imkovic, Phys. Rev. C 83, 034320 (2011).
\bibitem{PRC_75_051303_2007} 
M. Kortelainen, J. Suhonen, Phys. Rev C 75, 051303(R) (2007).
\bibitem{PRC_87_064302_2013}
M. T. Mustonen and J. Engel, Phys. Rev. C 87, 064302 (2013).
\bibitem{PRL_100_052503_2008} 
E. Caurier \textit{et al.}, Phys. Rev. Lett. 100, 052503 (2008). 
\bibitem{NPA_818_139_2009} 
J. Men\'endez \textit{et al.}, Nucl. Phys. A 818, 139 (2009). 
\bibitem{PRC_86_067304_2012}
A. Neacsu, S. Stoica, and M. Horoi, Phys. Rev. C 86, 067304 (2012).
\bibitem{PRC_79_044301_2009}
J. Barea and F. Iachello, Phys. Rev. C \textbf{79}, 044301 (2009).
\bibitem{PRC_87_014315_2013}
J. Barea, J. Kotila, and F. Iachello, Phys. Rev. C 87, 014315 (2013).
\bibitem{PRC_82_064310_2010} 
P. K. Rath \textit{et al.}, Phys. Rev. C 82, 064310 (2010). 
\bibitem{PRL_105_252503_2010}
T. R. Rodr\'iguez and G. Martinez-Pinedo, Phys. Rev. Lett. 105, 252503 (2010).
\bibitem{PPNP_66_436_2011} 
T. R. Rodr\'iguez and G. Martinez-Pinedo, Prog. Part. Nucl. Phys. 66, 436 (2011).
\bibitem{PLB_719_174_2013}
T. R. Rodr\'iguez and G. Martinez-Pinedo, Phys. Lett. B 719, 174 (2013).
\bibitem{JM}
J. Men\'endez, \textit{et al.}, 2008. arXiv:0809.2183.
\bibitem{NPA_428_23_1984}
J.~F.  Berger et \textit{al.}, Nucl. Phys. A \textbf{428}, 23 (1984).
\bibitem{RingSchuck} 
P. Ring, P. Schuck, \textit{The nuclear many body problem}, Springer-Verlag, Berlin, 1980. 
\bibitem{RMP_75_121_2003} 
M. Bender, P.-H. Heenen, P.-G. Reinhard, Rev. Mod. Phys. 75, 121 (2003). 
\bibitem{PLB_704_520_2011}
N. L\'opez-Vaquero, T. R. Rodr\'iguez and J. L. Egido, Phys. Lett. B 704, 520 (2011).
\bibitem{NPA_696_467_2001} 
M. Anguiano, J. L. Egido, L. M. Robledo, Nucl. Phys. A 696, 467 (2001). 
\bibitem{footnote} 
With a state independent pairing interaction (monopole pairing), one would generate wave functions with different pairing content just constraining on the energy gap parameter $\Delta$.  For finite range forces, this constraint is completely equivalent~\cite{PLB_704_520_2011} to the one on the operator $<(\Delta\hat{A})^2>^{1/2}$. To get acquainted with this quantity we mention that for $^{136}$Ba and for $\beta_2=0.15$, the constraint to $\delta =1$ 
provides a state with a pairing energy of $-2$ MeV, $\delta =3$ one with  $-14$ MeV and $\delta =5$ one with  $-49$ MeV.
\bibitem{CPC_36_1603_2012}
M. Wang et \textit{al.}, Chinese Phys. C 36, 1603 (2012).
\bibitem{ADNDT_87_185_2004}
I. Angeli, At. Data Nucl. Data Tables 87, 185 (2004).
\bibitem{PRL_103_012503_2009}
K. Yako, et \textit{al.}, Phys. Rev. Lett. 103, 012503 (2009).
\bibitem{PRC_40_540_1989}
R. Madey, et \textit{al.}, Phys. Rev. C 40, 540 (1989).
\bibitem{PRC_55_2802_1997}
R.L. Helmer, et \textit{al.}, Phys. Rev. C 55, 2802 (1997).   
\bibitem{PRC_78_041602_2008}
H. Dohmann, et \textit{al.}, Phys. Rev. C 78, 041602(R) (2008).
\bibitem{PRC_71_054313_2005}
S. Rakers, et \textit{al.}, Phys. Rev. C 71, 054313 (2005).
\bibitem{RMP_77_427_2005}
E. Caurier, et \textit{al.}, Rev. Mod. Phys. 77, 427 (2005).
\bibitem{EPJA_33_237_2007}
S. Hilaire and M. Girod, Eur. Phys. J. A 33, 237 (2007).
\bibitem{PRC_83_015502_2011}
F. \u{S}imkovic, R. Hod\'ak, A. Faessler, and P. Vogel, Phys. Rev. C 83, 015502 (2011).
\bibitem{PRC_87_041304_2013}
Y. Toh  et \textit{al.}, Phys. Rev. C 87, 041304(R) (2013).
\bibitem{PRD_65_092007_2002}
C. E. Aalseth \textit{et al.}, Phys. Rev. D 65, 092007 (2002).
\bibitem{inprep}
T. R. Rodr\'iguez, J. Men\'endez, G. Mart\'inez-Pinedo, A. Poves, in preparation.
\bibitem{PRC_87_045501_2013}
F. \u{S}imkovic, V. Rodin, A. Faessler, and P. Vogel, Phys. Rev. C 87, 045501 (2013).
\end{thebibliography}
\end{document}